\documentclass[prd,nofootinbib,superscriptaddress,eqsecnum,tightenlines,11pt]{revtex4}

\usepackage{amsfonts}
\usepackage{amsmath}
\usepackage{hyperref}
\usepackage{stmaryrd}
\usepackage{color}
\usepackage{graphics,graphicx}

\def\be{\begin{equation}}
\def\ee{\end{equation}}
\def\bes{\begin{equation*}}
\def\ees{\end{equation*}}
\def\ba{\begin{eqnarray}}
\def\ea{\end{eqnarray}}

\def\eps{\varepsilon}
\def\lp{l_\text{Pl}}

\def\de{\mathrm{d}}
\def\Pexp{\overrightarrow{\exp}}

\newcommand{\f}{\frac}

\newcommand{\SU}{\text{SU}}

\newcommand{\SL}{\text{SL}}
\newcommand{\su}{\mathfrak{su}}
\renewcommand{\sl}{\mathfrak{sl}}

\renewcommand\time{\,{\scriptstyle{\times}}\,}

\begin{document}

\title{Spectra of geometric operators in three-dimensional LQG:\\From discrete to continuous}

\author{Jibril Ben Achour}
\affiliation{Laboratoire APC -- Astroparticule et Cosmologie, Universit\'e Paris Diderot Paris 7, 75013 Paris, France}
\author{Marc Geiller}
\affiliation{Institute for Gravitation and the Cosmos \& Physics Department, Penn State, University Park, PA 16802, U.S.A.}
\author{Karim Noui}
\affiliation{Laboratoire de Math\'ematiques et Physique Th\'eorique, Universit\'e Fran\c cois Rabelais, Parc de Grandmont, 37200 Tours, France}
\affiliation{Laboratoire APC -- Astroparticule et Cosmologie, Universit\'e Paris Diderot Paris 7, 75013 Paris, France}
\author{Chao Yu}
\affiliation{\'ENS de Lyon, 46, all\'ee d'Italie, 69007 Lyon, France\bigskip}

\begin{abstract}
We study and compare the spectra of geometric operators (length and area) in the quantum kinematics of two formulations of three-dimensional Lorentzian loop quantum gravity. In the $\SU(2)$ Ashtekar-Barbero framework, the spectra are discrete and depend on the Barbero-Immirzi parameter $\gamma$ exactly like in the four-dimensional case. However, we show that when working with the self-dual variables and imposing the reality conditions the spectra become continuous and $\gamma$-independent.
\end{abstract}

\maketitle

\section{Introduction}

\noindent In a series of papers \cite{GN1,GN2,Imm3D}, we have recently introduced and studied a model of three-dimensional gravity that has proven to be very useful to understand some of the fundamental features of canonical loop quantum gravity and spin foam models. This new formulation of gravity in three dimensions is particularly interesting because it admits a Barbero-Immirzi parameter, and can be written as a Holst or a Plebanski action. As such, it can be used to study the imposition of the simplicity constraints in the construction of spin foam models, the role of $\gamma$ and of the $\SU(2)$ Ashtekar-Barbero formulation, and the relation with the self-dual variables. In \cite{Imm3D}, it has been shown that this Lorentzian three-dimensional model can be written at the Hamiltonian level in terms of $\SU(2)$ canonical variables (just like in the four-dimensional case), which leads to discrete and $\gamma$-dependent geometrical operators. However, it turns out to be possible to simplify the Hamiltonian constraint (and write it as a flatness constraint) by working with complex canonical variables, provided that one imposes simplicity-like conditions ensuring that the corresponding connection be $\su(1,1)$-valued. In terms of these new variables, the spectra of the geometric operators become continuous and $\gamma$-independent.

In four-dimensional loop quantum gravity, the discreteness of quantum geometry at the Planck scale, together with the existence of a non-zero minimal eigenvalue for the area operator, play a crucial role in the loop quantum cosmology scenario of singularity resolution \cite{LQC} and are essential for the recovery of the semi-classical thermodynamical properties of black holes \cite{BHentropy1,BHentropy2,BHentropy3,BHentropy4}. Independently of the question of wether or not the geometric operators will remain discrete and $\gamma$-dependent at the dynamical level, it has been argued that already at the kinematical level a different choice of connection could potentially lead to continuous and $\gamma$-independent spectra \cite{alexandrov5}. This observation has always remained at the formal level, since no Hilbert space is known for the Lorentz-covariant theory in terms of the shifted connection of Alexandrov. It is however known that this approach is closely related to the quantization in terms of the complex self-dual Ashtekar connection, a route which has been recently reconsidered in the light of results on black hole entropy \cite{FGNP,BST,pranzetti} and the asymptotics of spin foam models \cite{BN}. In the absence of a known way of constructing a full quantum theory with the self-dual variables, one is forced to investigate toy models and to focus on particular calculations such as the derivation of black hole entropy. The hope is that this will provide informations as to how the self-dual quantum theory should be built. In particular, a key open question concerns the role of the Barbero-Immirzi parameter and of the $\SU(2)$ variables. Are these fundamental variables of quantum gravity with $\gamma$ playing a true physical role, or simply a necessary regularization procedure needed in order to deal with the non-compactness of the gauge group of Lorentzian gravity? It seems to us that the latter point of view should be adopted. In this short paper, we give an example of the above-mentioned duality between the nature of the geometric spectra (discrete or continuous) and the choice of connection. For this, we use the Holst formulation of three-dimensional gravity to compare the quantum theories built with the $\su(2)$ Ashtebar-Barbero and the $\sl(2,\mathbb{C})$ self-dual connection.

More precisely, our starting point for this study is the observation that the three-dimensional model of interest can be obtained from a symmetry reduction of the four-dimensional Holst action, defined by imposing a symmetry along an arbitrary spatial direction. Fortunately, this reduction preserves the presence of $\gamma$, and the resulting theory shares most of the properties of its four-dimensional counterpart. In particular, in the time gauge, the classical dynamical variables are given by an $\SU(2)$ Ashtekar-Barbero connection $A$ canonically conjugated to an $\su(2)$-valued electric field $E$. Furthermore, the six first class constraints that generate the isometries of the locally flat space-time can be decomposed, as in four dimensions, in terms of a Gauss constraint, a vectorial constraint, and a scalar constraint. Since the theory is topological in three dimensions, the physical degrees of freedom are captured by the holonomies $h_\ell(A)$ of the connection along the links $\ell$ of a unique graph $\Gamma$ (sufficiently refined to resolve the topology of the spatial surface), and the ``one-dimensional fluxes'' of the electric field $X_\ell$ along links $\ell^*$ (dual to $\ell$) of the dual graph $\Gamma^*$. At the quantum level, these (non-local) variables are promoted to non-commutative operators which satisfy the three-dimensional analogue of the holonomy-flux algebra. As usual, kinematical states are $\SU(2)$ spin networks associated to the graph $\Gamma$, and the geometric operators are constructed in terms of the flux operators $X_\ell$. The spectrum of the length operator is given by a sum of fundamental contributions $\ell(j)=\gamma\lp\sqrt{j(j+1)}$, where $j$ labels unitary irreducible representations of $\SU(2)$, and $\lp$ is the three-dimensional Planck length. Motivated by the fact that this $\SU(2)$ Hamiltonian theory should describe Lorentzian three-dimensional gravity, i.e. $\SU(1,1)$ BF theory, the Hamiltonian constraint can be recast into the form of a flatness constraint. The canonical connection then becomes complex and $\sl(2,\mathbb{C})$-valued, although it is defined with $\gamma\in\mathbb{R}$. On top of this complex formulation, it is possible to impose two types of linear simplicity conditions similar to that used in spin foam models. The first form of the constraint (which relates the $\su(2)$ generators and their complement in $\sl(2,\mathbb{C})$) gives back the $\su(2)$-valued Ashtekar-Barbero connection, while the second one (which relates the $\su(1,1)$ generators and their complement in $\sl(2,\mathbb{C})$) leads to an $\su(1,1)$ connection. Constructing the quantum theory with the latter, we observe that the length operator is given by a sum of fundamental contributions $\ell(s)=\lp\sqrt{s^2+c}$ (with $c=1/4$ or $c=0$ depending on the regularization of the length operator), where $s$ is a real parameter labeling the continuous series of representations of $\SU(1,1)$. This spectrum is therefore continuous and $\gamma$-independent.

This paper is organized as follows. In the next section, we briefly review the classical formulation that serves as the starting point for this analysis. In section \ref{sec:3}, we present the two different choices of connection that can be made at the classical level, and discuss the resulting quantum theories. Section \ref{sec:4} is devoted to the introduction of the classical geometric operators, while their spectra are studied in section \ref{sec:5}.

\section{Classical theory}

\noindent Let us start with the four-dimensional Holst action
\be
S_\text{4D}[e,\omega]=\int_{\mathcal{M}_4}\left(\frac{1}{2}\eps_{IJKL}e^I\wedge e^J \wedge 
F^{KL}+\frac{1}{\gamma}\delta_{IJKL}e^I\wedge e^J\wedge F^{KL}\right),
\ee
and perform a spacetime symmetry reduction that keeps the internal gauge group $\SL(2,\mathbb{C})$ unbroken. For this, we assume that the four-dimensional spacetime manifold has the topology $\mathcal{M}_4=\mathcal{M}_3\times\mathbb{S}^1$, where $\mathcal{M}_3$ is a three-dimensional spacetime manifold, and $\mathbb{S}^1$ is a spatial direction along which the theory is invariant. Denoting by $x_3$ the coordinate along $\mathbb{S}^1$, the invariance of the basic fields can be expressed as
\be\label{symmetry conditions}
\partial_3=0,\qquad\qquad\omega_3^{IJ}=0.
\ee
As a consequence, the resulting reduced action is given by
\be\label{3D action}
S[e,x,\omega]=\int_{\mathcal{M}_3}\de^3x\,\eps^{\mu\nu\rho}\left(\frac{1}{2}\eps_{IJKL}x^Ie_\mu^JF_{\nu \rho}^{KL}+\frac{1}{\gamma}\delta_{IJKL}x^Ie_\mu^JF_{\nu \rho}^{KL}\right),
\ee
where $\omega^{IJ}_\mu$ is an $\sl(2,\mathbb{C})$-valued connection over $\mathcal{M}_3$, $e^I_\mu$ is an $\sl(2,\mathbb{C})$-valued one form over $\mathcal{M}_3$, and $x^I$ is an $\sl(2,\mathbb{C})$-valued scalar on $\mathcal{M}_3$. As usual, $\eps^{\mu\nu\rho}$ is the totally antisymmetric spacetime tensor, $\eps_{IJKL}$ denotes the totally antisymmetric internal tensor, and $\delta_{IJKL}=(\eta_{IK}\eta_{JL}-\eta_{IL}\eta_{JK})/2$ is defined in term of the flat Minkowski metric
$\eta$. We assume that $\mathcal{M}_3=\Sigma_2\times\mathbb{R}$, where $\Sigma_2$ is a two-dimensional surface with no boundaries.

In order to mimic the construction of four-dimensional loop quantum gravity, we performed in \cite{Imm3D} the canonical analysis of the three-dimensional action (\ref{3D action}) in the time gauge. This gauge fixing is defined by the conditions $x^0=e^0_a=0$, and breaks the internal gauge group $\SL(2,\mathbb{C})$ into its maximal compact subgroup $\SU(2)$. Just like in the four-dimensional theory, the resulting phase space is parametrized by the canonical pair $(E^a_i(x),A_b^j(y))$, where $A^i_a$ is the three-dimensional analogue of the $\su(2)$ Ashtekar-Barbero connection, and $E^a_i$ its conjugate electric field. These variables satisfy the Poisson bracket
\be\label{Poisson bracket}
\{E^a_i(x),A_b^j(y)\}=\gamma\delta^a_b\delta^j_i\delta^2(x-y),
\ee
and are subject to the following Gauss, vectorial, and scalar constraints:
\be\label{constraints}
G=\partial_aE^a+A_a\time E^a,\qquad
H_a=\eps_{ab}E^b\cdot F_{12},\qquad
H_0=x\cdot\left(F_{12}-(1+\gamma^{-2})K_1\time K_2\right),
\ee
where $F_{12}=\partial_1A_2-\partial_2A_1+A_1\time A_2$ is the curvature of $A$. The variable $\gamma^{-1}K^i_a$ is the three-dimensional analogue of the extrinsic curvature, and can be written as $\gamma^{-1}K^i_a=\gamma^{-1}\big(A^i_a+\omega^i_a(E)\big)=\omega_a^{0i}$, where $\omega(E)$ is the reduced version of the four-dimensional Levi-Civita connection $\Gamma(E)$. Its explicit expression is given by
\be
\omega_a(E)=u\time\partial_au+\eps_{ab}\frac{E^b\cdot\partial_cE^c}{|E^1\time E^2|}u,
\ee
where $u^i=x^i/\sqrt{x^2}$. To finish with this brief overview of the phase space structure, let us mention that $x$ is colinear to $E^1\time E^2$, which makes the above expression for $H_0$ strictly similar to the scalar constraint of the four-dimensional full theory. Furthermore, because the vectors $K_a$ are orthogonal to $u$ \cite{Imm3D}, the quantity $K_1\time K_2$ is in the direction of $u$, and $H_a$ and $H_0$ can be viewed as the component of the same vector $H$ defined by
\be\label{F=0 in time gauge}
H={F}_{12}-(1+\gamma^{-2})K_1\times K_2.
\ee
This property is in some sense responsible for the fact that this three-dimensional model is exactly soluble.

\section{Quantum theory: from $\boldsymbol{\SU(2)}$ to $\boldsymbol{\SU(1,1)}$ via $\boldsymbol{\SL(2,\mathbb{C})}$}
\label{sec:3}

\subsection{$\boldsymbol{\SU(2)}$ quantum theory}

\noindent We now have all the ingredients to perform the loop quantization of the theory. Kinematical states are cylindrical functions of the connection associated with graphs embedded in the spatial surface $\Sigma_2$. Because of the topological nature of three-dimensional gravity, a single graph $\Gamma$ is sufficient to capture all the physical content of the theory. On this graph, let us introduce the holonomies of the connection along the links $\ell\in\Gamma$, and the fluxes of the electric field along links $\ell^*\in\Gamma^*$ dual to $\ell$. These variables, i.e.
\begin{eqnarray}\label{holflux}
\SU(2)\ni h_\ell(A)=\Pexp\int_\ell A_a^i\de x^a\tau_i,\qquad\qquad
\su(2)\ni X^i_{\ell^*}=\int_{\ell^*}\eps_{ab}E^a_i\de x^b,
\end{eqnarray}
are the building blocks of the three-dimensional (classical and quantum) holonomy-flux algebra. The action of the flux operator on the holonomies evaluated in the spin-$j$ representation of $\su(2)$ is given by
\be\label{actionofL}
X^i_{\ell^*}\triangleright\mathbf{D}^{(j)}(h_{\ell'}(A))=\mathrm{i}\epsilon\gamma\lp\delta_{\ell,\ell'}\mathbf{D}^{(j)}(h_{\ell<c}(A))J_i\mathbf{D}^{(j)}(h_{\ell>c}(A)),
\ee
where $\epsilon\in\{-1,+1\}$ is the index between $\ell$ and $\ell^*$, the point $c$ denotes the intersection $\ell\cap\ell^*$, and $\mathbf{D}^{(j)}:\SU(2) \longrightarrow\mathbb{V}^{(j)}$ denotes the spin-$j$ representation matrix associated to the $(2j+1)$-dimensional vector space $\mathbb{V}^{(j)}$. The elements $J_i$ generate the Lie algebra $\su(2)$ and satisfy, by convention, the Lie algebra commutation relations $[J_i,J_j]=\eps_{ij}^{~~k} J_k$, where $\eps_{ijk}$ is the totally antisymmetric tensor with $\eps_{123}=1$, and indices are lowered and raised with the flat Euclidean metric $\delta_{ij}$. As usual, cylindrical functions associated to the graph $\Gamma$ form the kinematical Hilbert space where the scalar product is constructed from the $\SU(2)$ Haar measure.

\subsection{Obtaining the $\boldsymbol{\SL(2,\mathbb{C})}$ connection}

\noindent In principle, physical states and observables should be constructed by solving the quantum constraint $H\simeq0$. In this set of constraints, only the imposition of the scalar constraint $H_0\simeq0$ is problematic since the vectorial constraint $H_a\simeq0$ imposing spatial diffeomorphism invariance has already been solved implicitly by fixing the graph $\Gamma$. The difficulty in solving $H_0\simeq0$ is essentially the same as in the four-dimensional theory, and requires an appropriate regularization of the term $K_1\time K_2$ when written in terms of the non-linear expression $\omega(E)$. This difficulty can however be bypassed by noticing that in this three-dimensional model the constraints $H\simeq0$ given by (\ref{F=0 in time gauge}) can be written in the form of the flatness constraint of a BF theory. More precisely, one can look for a connection $\mathbf{A}$ for which $H\simeq0$ is equivalent to  $\mathcal{F}_{12}\simeq 0$, where $\mathcal{F}$ is the curvature of $\mathbf{A}$. It was shown in \cite{Imm3D} that, up to gauge transformations, there are only two connections that satisfy this requirement. These are the self-dual and anti self-dual components of the initial $\SL(2,\mathbb{C})$ connection $\omega^{IJ}_a$. The connection $\mathbf{A}$ is therefore complex and has to be interpreted as an $\sl(2,\mathbb{C})$-valued connection. Therefore, we have somehow traded the problem of dealing with the complicated Hamiltonian constraint (\ref{F=0 in time gauge}) for that of imposing the reality conditions. The same fact is obviously true in four dimensions, where one can take $\gamma=\mathrm{i}$ to obtain the self-dual scalar constraint, at the expense of working with the complex Ashtekar connection.

We arrive at the conclusion that the quantum dynamics can in principle be solved in terms of the connection $\mathbf{A}$, since with this variable the Hamiltonian constraint is manageable. This means that one should consider $\SL(2,\mathbb{C})$ spin network states instead of the standard $\SU(2)$ ones, which introduces numerous technical problems due to the non-compactness of the Lorentz group. A priori, such spin network states require a regularization in order to be well-defined. Fortunately, as we are about to see, this problem does drastically simplify due to the constraints satisfied by the complex connection $\mathbf{A}$.

\subsection{Simplicity constraints and the $\boldsymbol{\SU(1,1)}$ connection}

\noindent As explained in \cite{Imm3D}, the connection $\mathbf{A}$ can be written in the form
\be\label{decomposition of A}
{\mathbf{A}}=\left[-u\time\de u\cdot(J\pm\gamma^{-1}P)\right]+
\left[(A\cdot u)(u\cdot J)\mp\gamma^{-1}(A\time u)\cdot(P\time u)\right],
\ee
where $J_i$ are the $\su(2)$ generators introduces above, and $P_i$ are the boost generators satisfying the standard $\sl(2,\mathbb{C})$ commutation relations. One can see from this expression that $\mathbf{A}$ possesses two different parts, which are the two terms between the square brackets. Interestingly, while the first term has no direct algebraic interpretation, the second term defines an $\su(1,1)$ component. Indeed, it can easily be shown that the elements $J\cdot u$ and the two independent components of $P\time u$ generate the Lie algebra $\su(1,1)$.

Now, one can use the fact that there exists another legitimate choice of gauge for the action (\ref{3D action}). Indeed, in the gauge $x^I=(0,0,0,1)$, the action (\ref{3D action}) reduces to that of $\SU(1,1)$ BF theory, which is in complete agreement with the fact that it describes Lorentzian three-dimensional gravity. For this reason, consistency of the theory in the time gauge requires that we impose that the connection $\mathbf{A}$ be $\su(1,1)$-valued as well. This requirement can be met if the first term in (\ref{decomposition of A}) is vanishing, which leads to the constraint $(J\pm\gamma^{-1}P)\time u=0$. This simplicity-like constraint is equivalent to a reality condition, in the sense that it selects a real section of $\SL(2,\mathbb{C})$ corresponding to its non-compact subgroup $\SU(1,1)$. Once it is imposed at the classical level, the linear simplicity constraint reduces the connection (\ref{decomposition of A}) to the following $\su(1,1)$ connection:
\be\label{cal A}
\mathbf{A}=\mp\mathrm{i}\gamma^{-1}(A^1F_1+A^2F_2-\mathrm{i}A^3F_0),
\ee
where $A^i$ are the components of the initial Ashtekar-Barbero connection $A$ expressed in the basis $(J_1,J_2,J_3)$, and the family $(F_0,F_1,F_2)$ generates $\su(1,1)$ with the commutation relations
\be
[F_1,F_2]=\mathrm{i}F_0,\qquad
[F_0,F_2]=\mathrm{i}F_1,\qquad
[F_0,F_1]=-\mathrm{i}F_2.
\ee
These generators are related by $(F_0,F_1,F_2)=-(\mathrm{i}J_3,J_1,J_2)$.

Now that the $\su(1,1)$ connection (\ref{cal A}) is defined, one can in principle construct the physical Hilbert space by imposing the quantum $\SU(1,1)$ flatness constraint. Physical states are $\SU(1,1)$ spin networks with support on the graph $\Gamma$. This defines the vector space structure of the physical Hilbert space, and it remains to define the scalar product between the states and eventually to eliminate the zero-norm states. In order to do so, it is much simpler to choose $\Gamma$ to be a minimal graph. When the spatial slice $\Sigma_2$ is a Riemann surface of genus $g$ with no boundaries, the minimal graph consists in only one vertex $v$ and $2g$ loops $(a_1,b_1,\dots,a_g,b_g)$ starting and ending at $v$. The loops $(a_i,b_i)$, with $i\in\llbracket1,g\rrbracket$, are the non-contractible loops around the handles of $\Sigma_2$, and they can be identified with the standard generators of the fundamental group $\pi^1(\Sigma_2)$ of the surface. For obvious reasons, such a minimal graph is called a flower graph. Physical states are then totally defined by complex-valued functions $\varphi$ on $\mathcal{G}_0=\SU(1,1)^{\otimes2g}/\SU(1,1)$, where the coset by $\SU(1,1)$ traduces the gauge-invariance at the unique vertex $v$ of $\Gamma$, and the physical inner product between two such states is formally defined by
\be\label{physical product}
\langle\varphi_1,\varphi_2\rangle=\int\de\mu(a,b)\overline{\varphi_1(a,b)}\delta([a,b])\varphi_2(a,b),
\ee
where $\de\mu(a,b)$ is a ``regularized'' measure on $\mathcal{G}_0$, $\delta$ is the Dirac distribution on $\SU(1,1)$, and the commutator is $[a,b]=\prod_{i=1}^g a_ib_ia_i^{-1}b_i^{-1}$. For simplicity, we have identified here the states with the corresponding funtions on $\mathcal{G}_0$. At this point, the physical scalar product and the physical Hilbert space are defined only at the formal level, but we will make this construction concrete and explicit later on in the case of the torus. To finish, let us recall that the physical scalar product (\ref{physical product}) should reproduce the Lorentzian Ponzano-Regge amplitudes \cite{Freidel}, just like it is the case for the Euclidean signature \cite{NP1,NP2}.

\section{Classical geometric operators}
\label{sec:4}

\noindent We now introduce the three-dimensional length and area operators. For this, we first define the four-dimensional area and volume operators, and then reduce them with the symmetry (\ref{symmetry conditions}). Recall that in four spacetime dimensions the Hamiltonian theory is written assuming that $\mathcal{M}_4=\Sigma_3\times\mathbb{R}$. The area $\alpha_4(S)$ and volume $\nu_4(R)$ operators are then defined on the spatial slice $\Sigma_3$, and measure the area and the volume of a surface $S\subset\Sigma_3$ and a region $R\subset\Sigma_3$, respectively. The classical expressions for the area $\alpha_4(S)$ of a surface $S$ of normal $n$, and the volume $\nu_4(R)$ of a bounded region $R$, depend only on the electric field $E^\mu=\eps^{\mu\nu\rho}e_\nu\time e_\rho/2$, with $\mu\in\{1,2,3\}$, and are given by
\be\label{A and V in 4D}
\alpha_4(S)=\int_S\de^2x\,\sqrt{E^\mu\cdot E^\nu n_\mu n_\nu},\qquad\qquad
\nu_4(R)=\int_R\de^3x\,\sqrt{\left|\frac{1}{3!}\eps_{\mu\nu\rho}E^\mu\cdot E^\nu\time E^\rho\right|}.
\ee
Notice that for $\mu=a\in\{1,2\}$, $E^\mu$ coincides with the electric field (\ref{Poisson bracket}) of the three-dimensional action (\ref{3D action}), and is given by $E^a=\eps^{ab}e_b\time x$, whereas for $\mu=3$ we have $E^3=e_1\time e_2$.

\begin{figure}[h]
\includegraphics[scale=0.45]{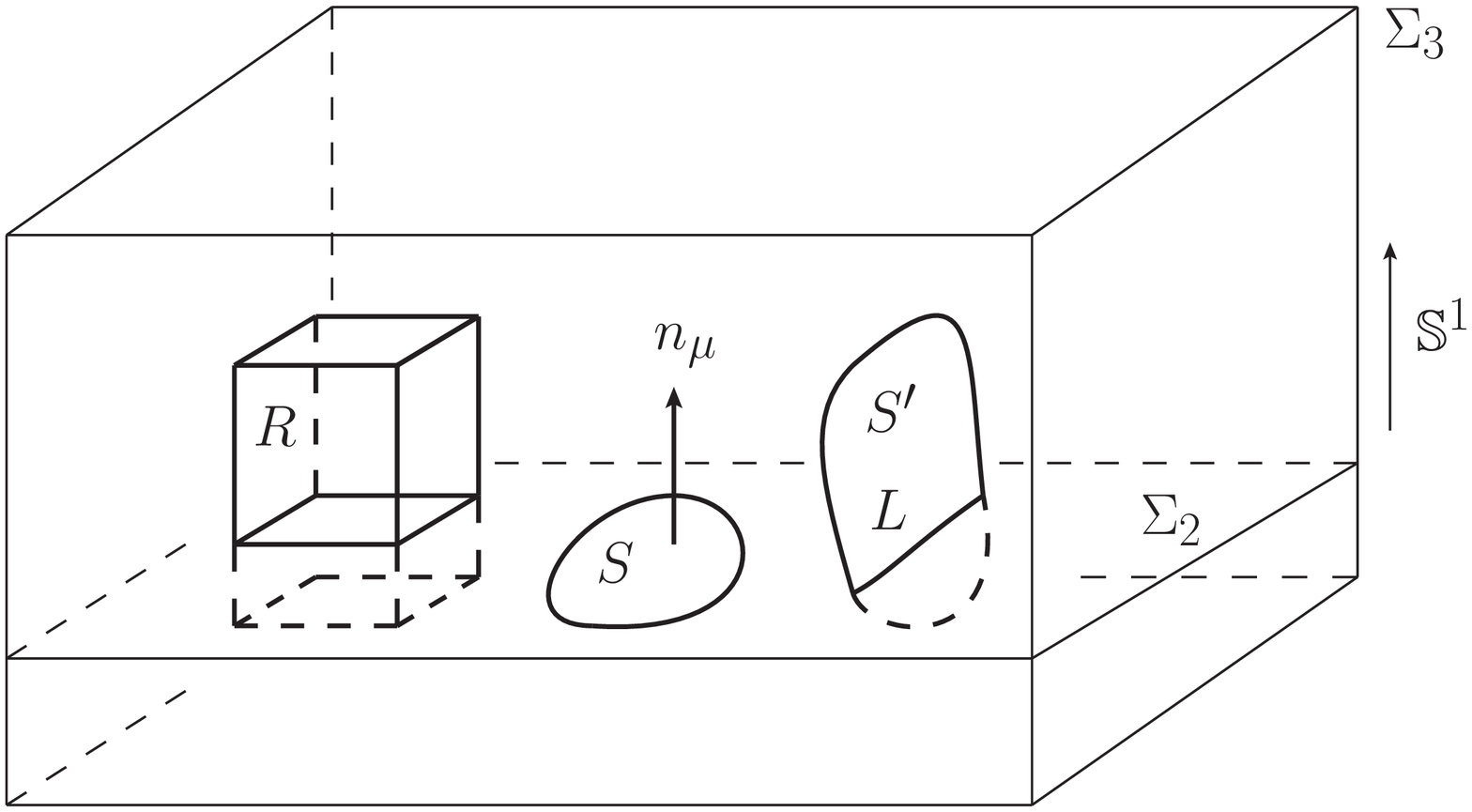}
\caption{Decomposition $\Sigma_3=\Sigma_2\times\mathbb{S}^1$ and representation of the two types of surfaces that can be embedded in $\Sigma_3$. The surface $S\subset\Sigma_2$ is a two-dimensional submanifold of $\Sigma_2$, while the surface $S'$ is of the type $S'=L\times I$, where $L$ is a one-dimensional submanifold of $\Sigma_2$ and $I$ a segment in the direction $\mu=3$. The length operator on $\Sigma_2$ can be derived from the area operator $\alpha_4(S')$ on $\Sigma_3$, while the area operator on $\Sigma_2$ can be derived from the volume operator $\nu_4(R)$ on $\Sigma_3$.}
\label{operators}
\end{figure}

The compactification that we have used to obtain the three-dimensional Holst action is of the type $\mathcal{M}_4=\mathcal{M}_3\times\mathbb{S}^1$. Since we have further assumed that $\mathcal{M}_3=\Sigma_2\times\mathbb{R}$ in order to perform the canonical analysis, we can write that
\be
\mathcal{M}_4=\Sigma_3\times\mathbb{R}=\mathcal{M}_3\times\mathbb{S}^1=\Sigma_2\times\mathbb{R}\times\mathbb{S}^1.
\ee
In other words, the spatial slice $\Sigma_3$ on which the operators (\ref{A and V in 4D}) are defined is given by $\Sigma_3=\Sigma_2\times\mathbb{S}^1$. Therefore, the surface $S\subset\Sigma_3$ measured by $\alpha_4(S)$ can be of two types, as represented in figure \ref{operators}. It can either be of the form $S=L\times I$, where $L$ is a path (i.e. a one-dimensional submanifold) in $\Sigma_2$ and $I$ a segment in $\mathbb{S}^1$, or it can be a two-dimensional submanifold of $\Sigma_2$.
Let us start with the case $S=L\times I$, where $L$ is a path in $\Sigma_2$ and $I$ a segment of length $d$ in the third direction $\mu=3$. Then, the normal $n_\mu$ is orthogonal to the third direction and has only components $(n_1,n_2)$ in the directions $a=1$ and $a=2$ (i.e. in $\Sigma_2$). The area reduces to $\alpha_4(S)=d\lambda(L)$, where $\lambda(L)$ is the length of $L$ given by
\be\label{3D length}
\lambda(L)=\frac{1}{d}\alpha_4(S)=\int_L\de x\,\sqrt{E^a\cdot E^bn_an_b}. 
\ee
In this way, we obtain immediately the length of any curve $L$ in $\Sigma_2$ in terms of the electric field and the unit vector $n$ normal to this curve.

When the surface $S$ is a two-dimensional submanifold of $\Sigma_2$, then it is immediate to see that $\alpha_4(S)$ gives the area of the surface $S$. In this case, $n$ is necessarily in the third direction, i.e. $n_1=n_2=0$ and $n_3=1$, and the expression $\alpha_4(S)$ for the area reduces to
\be
\alpha(S)=\int_S\de^2x\,\sqrt{E^3\cdot E^3},
\ee
which can be expressed in terms of the triad field since $E^3=e_1\time e_2$. To write this expression in terms of the electric field $E^a$ only, we use a symmetry of the action (\ref{3D action}) described in \cite{Imm3D}, which states that $x$ can be chosen to be of unit norm (${x^2}=1$), and that it is orthogonal to $e^a$ (i.e. $e_a\cdot x=0$). If we make this choice, then the area takes the simple form
\be\label{3D area}
\alpha(S)=\int_S\de^2x\,\sqrt{|E^1\time E^2|^2}=\int_S\de^2x\,\sqrt{(E^1)^2 (E^2)^2 - (E^1 \cdot E^2)^2}.
\ee
The operators $\lambda(L)$ and $\alpha(S)$ defined in (\ref{3D length}) and (\ref{3D area}) measure respectively the length of a curve $L\subset\Sigma_2$ and the area of a surface $S\subset\Sigma_2$ in our three-dimensional model.

Notice that we could have derived the expression for the area operator $\alpha(S)$ of the three-dimensional theory by considering the volume operator $\nu_4(R)$ defined for a region $R$ of the type $R=S\times I$, where $S$ is a surface in $\Sigma_2$ and $I$ a segment of length $d$ in $\mathbb{S}^1$ (see figure \ref{operators}). Indeed, for such a region we have $\nu_4(R)=d\alpha(S)$, and one can then directly obtain (\ref{3D area}).

Now that we have derived the classical expressions for the geometric operators of the three-dimensional theory (that are defined on the two-dimensional surface $\Sigma_2$), we can quantize them and study their action on the above-defined $\SU(2)$ and $\SU(1,1)$ states. As it is the case for the four-dimensional volume operator $\nu_4(R)$, the three-dimensional area operator $\alpha(S)$ requires a regularization in order to be well-defined in terms of the quantum flux operators $X_{\ell^*}$ \cite{rovelli-smolin-area-volume,ashtekar-lewandowski-volume,giesel-thiemann-vol1,giesel-thiemann-vol2,thiemann-brunnemann,brunnemann-rideout1,brunnemann-rideout2}. These regularizations have been studied and well-understood in the literature, and they will not affect our result.

\section{Quantum geometric operators}
\label{sec:5}

\noindent The length operator is associated to links $\ell^*$ of the dual graph $\Gamma^*$, and the length $\lambda(\ell^*)$ of the link $\ell^*$ dual to $\ell$ can be expressed in terms of the fluxes (\ref{holflux}) according to
\be\label{length operator}
\lambda(\ell^*)=\sqrt{X_{\ell^*}^2}.
\ee
The length $\lambda(L^*)$ of any path $L^*\in\Gamma^*$ defined by a composition $L^*=\ell_1^* \circ\dots\circ\ell_n^*$ of a sequence of $n$ elementary links\footnote{A physical observable is given by a closed loop $L$ around non-contractible handles of the surface $\Sigma_2$. An open link $L$ would not correspond to a physical (Dirac) observable apart if particles are coupled to gravity and located at the end points of $L$ .} is simply given by the sum $\sum_{i=1}^n \lambda({\ell_i^*})$. Therefore, computing the action of $\lambda(\ell^*)$ on the holonomy $h_{\ell'}(A)$ or $g_{\ell'}(\mathbf{A})$ allows to compute the action of any length operator $\lambda(L^*)$ on the states of the $\SU(2)$ or $\SU(1,1)$ quantum theories. This action is non-trivial only when $\ell=\ell'$, it is diagonal and given by
\ba\label{lengthspectrum}
\lambda(\ell^*)\triangleright\mathbf{D}^{(j)}(h_\ell(A))&=&\gamma\sqrt{C(\su(2))}\mathbf{D}^{(j)}(h_\ell(A)),\\
\lambda(\ell^*)\triangleright\mathbf{D}^{(s)}(g_\ell(\mathbf{A}))&=&\sqrt{C(\su(1,1))}\mathbf{D}^{(s)}(g_\ell(\mathbf{A})),
\ea
where $C(\su(2))=-J^2$ and $C(\su(1,1))=F_1^2+F_2^2-F_0^2$ are the quadratic Casimir operators of the Lie algebras $\su(2)$ and $\su(1,1)$, respectively. We have used the notation $j$ to denote unitary irreducible representations of $\su(2)$ and generically $s$ for those of $\su(1,1)$ (whose class has not been fixed yet). While $C(\su(2))=j(j+1)$ is always positive and discrete because $j$ is a non-negative half-integer, the sign of $C(\su(1,1))$ depends on the class of $\su(1,1)$ representations $s$ under consideration. In order to have a self-adjoint length operator (with real and positive eigenvalues), $s$ must belong to the continuous series. In this case, the spectrum $\sqrt{C(\su(1,1))}=\sqrt{s^2+1/4}$ of the length operator is continuous. Before studying the area operator, let us make some important remarks. 

\begin{enumerate}
\item First, we would like to emphasize that, in addition to being continuous, the length spectrum becomes also independent of the Barbero-Immirzi parameter $\gamma$. The reason is simple and comes from the overall factor of $\gamma^{-1}$ in the expression (\ref{cal A}) for $\mathbf{A}$, which cancels with the factor of $\gamma$ coming from the Poisson bracket (\ref{Poisson bracket}) when computing the action of $X_{\ell^*}$ on $g_\ell(\mathbf{A})$, i.e.\footnote{The basis $F_i$ as to be understood with $i\in\{0,1,2\}$, while for $J_i$ we have $i\in\{1,2,3\}$.}
\be
X^i_{\ell^*}\triangleright\mathbf{D}^{(s)}(g_{\ell'}(\mathbf{A}))
=(-\mathrm{i}\gamma)(\mp\mathrm{i}\gamma^{-1})\lp\delta_{\ell,\ell'}\mathbf{D}^{(s)}(g_{\ell<c}(\mathbf{A}))(\sigma F_i)\mathbf{D}^{(s)}(g_{\ell>c}(\mathbf{A})),
\ee
where $\sigma\in\{1,-\mathrm{i}\}$ is equal to $-\mathrm{i}$ when one acts with the third component $X^3_{\ell^*}$ of the flux.

\item The factor of $\sigma$ is responsible for the fact that the action of the $\su(2)$-invariant operator $X^2_{\ell^*}=X^i_{\ell^*}\delta_{ij}X^j_{\ell^*}$ produces the $\su(1,1)$ quadratic Casimir when it acts on the holonomies $g_\ell(\mathbf{A})$. Of course, one could think of writing the action of the length operator on the $\SU(2)$ holonomies $h_\ell(A)$ in the basis $F_i$ instead of $J_i$, but then the spectrum would come with an incorrect minus sign.

\item Even if the spectrum (\ref{lengthspectrum}) is continuous when using the connection $\mathbf{A}$ instead of $A$, there exists a length gap between ``no length'' and the first non-trivial possible length. Another regularization, discussed in \cite{freidel-livine-rovelli}, would give $s$ instead of $\sqrt{s^2+1/4}$ for the action of the length operator on holonomies colored with a representation $s$, and no length gap would exist in this case. We refer the reader to \cite{freidel-livine-rovelli} for a discussion of this ambiguity.
\end{enumerate}

The area operator in three dimensions is the analogue of the volume operator in four dimensions, and it acts on the vertices of the spin networks. For obvious reasons, we will be interested only in its action on three-valent vertices. Denoting by $\ell_1$, $\ell_2$ and $\ell_3$ the three links meeting at the vertex $v$, we will use the notation $\alpha(\ell_1,\ell_2,\ell_3)$ for the area operator at $v$. It is straightforward to obtain the following expression for the regularized area operator acting on the three-valent vertex $v$ \cite{freidel-livine-rovelli}:
\be
\alpha(\ell_1,\ell_2,\ell_3)^2=\f{1}{6}\sum_{I<J}\left(X_{\ell^*_I}^2X_{\ell^*_J}^2-(X_{\ell^*_I}\cdot X_{\ell^*_J})^2 \right),
\ee
where $1\leq I<J\leq3$ label the edges meeting at the vertex. One can then show that the action of the area operator is diagonal, with eigenvalues given by
\be
\alpha(\ell_1,\ell_2,\ell_3)^2=\f{1}{6}\sum_{(I,J)}\left(\lambda_I^2\lambda_J^2-(\xi_I\cdot\xi_J)^2\right),
\ee
where the sum contains only the three terms $(1,2)$, $(2,3)$ and $(1,3)$. The Lie algebra elements $\xi$ denote the $\su(2)$ or $\su(1,1)$ generators depending on the states that we act on (i.e. the $\SU(2)$ or $\SU(1,1)$ spin networks), and the scalar product in the internal space is defined with the flat Euclidean metric $\delta_{ij}$ or the flat Minkowskian metric $\eta_{ij}$, depending on the choice of connection. Finally, $\lambda_I$ denotes the eigenvalue of the length operator $\lambda({\ell^*_I})$ (\ref{length operator}). Using the Gauss law at the vertex $v$ to write $X_{\ell^*_1}^i+X_{\ell^*_2}^i+X_{\ell^*_3}^i=0$, an immediate computation leads to
\ba\label{areaproduct}
\alpha(\ell_1,\ell_2,\ell_3)^2&=&\f{1}{6}\sum_{(I,J)}\left(\lambda_I^2\lambda_J^2 
-\f{1}{4}(\lambda_I^2+\lambda_J^2-\lambda_K^2)^2\right)\nonumber\\
&=&\f{1}{16}(\lambda_1+\lambda_2+\lambda_3)(-\lambda_1+\lambda_2+\lambda_3)
(\lambda_1-\lambda_2+\lambda_3)(\lambda_1+\lambda_2-\lambda_3),
\ea
where $K \neq I,J$. We obtain the classical expression for the area of a triangle with edges of length $\lambda_I$, and the geometrical interpretation of this formula is therefore immediate. However, an extra condition has to be imposed in order for the area eigenvalues to be real. Indeed, the product (\ref{areaproduct}) is positive only if the three lengths $\lambda_I$ satisfy the triangular inequalities. Contrary to the $\SU(2)$ kinematical case where the triangular inequalities are automatically satisfied due to the standard properties of tensor products between unitary irreducible representations, here this condition has to be imposed by hand because the representations are in the continuous series of $\SU(1,1)$.

\begin{figure}[h]
\includegraphics[scale=0.45]{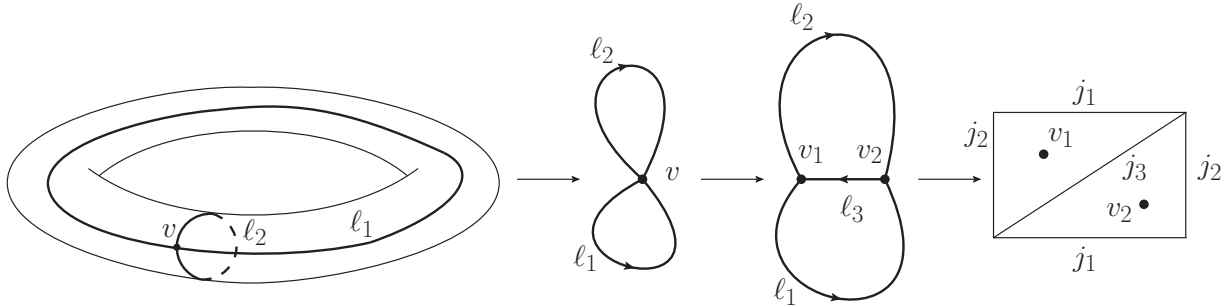}
\caption{Discretization of a torus by a minimal flower graph. The flower consists only in a single four-valent vertex $v$ and two loops, $\ell_1$ and $\ell_2$, corresponding to the non-contractible cycles of the torus. A spin network state associated to the flower graph is an assignment of unitary irreducible representations of the suitable Lie algebra to the two loops and to the vertex. The vertex being four-valent, it can be decomposed into two three-valent vertices, with the spin previously attached to $v$ now coloring the link $\ell_3$. As shown on the right, this graph with three links can be seen as dual to the rectangle with identified opposite edges that defines the torus. This rectangle can be triangulated into two triangles sharing the side of length $j_3$ and dual to the vertices $v_1$ and $v_2$. The spins $j_1$ and $j_2$ coloring the loops $\ell_1$ and $\ell_2$ represent the edge lengths of the rectangle.}
\label{torus}
\end{figure}

To finish, let us study more precisely the case in which $\Sigma_2$ is a torus. The torus can be represented by the flower graph with two petals as shown in figure \ref{torus}.
In this case, one can construct the kinematical spin network states of the $\SU(2)$ theory based on the three-dimensional Ashtekar-Barbero connection $A^i_a$. These states, denoted by $\varphi_\text{kin}(j_1,j_2,j_3)$, are totally characterized by the three spin labels $(j_1,j_2,j_3)$ coloring the two loops $\ell_1$ and $\ell_2$ (for the first two labels) and the vertex $v$. Note that the choice of a ``shape'' for the intertwiner is not important (i.e. the choice of a triangulation for the rectangle in figure \ref{torus} is arbitrary). Our choice, which is depicted in figure \ref{torus}, corresponds to a triangulation of the parallelogram into two triangles. Since the graph is minimal, the kinematical scalar product is given by
\be
\langle\varphi_\text{kin}(j_1,j_2,j_3),\varphi_\text{kin}(j'_1,j'_2,j'_3)\rangle=\prod_{i=1}^3\delta_{j_i,j_i'}, 
\ee
where $\delta_{j,j'}$ denotes the Kronecker delta. The length and area operators of the $\SU(2)$ kinematical theory are unitary with respect to this scalar product since their action is diagonal with positive eigenvalues. Furthermore, the geometric interpretation of the spin labels is immediate: they are associated with the length of the three edge of the (similar) triangles that discretize the parallelogram whose opposite sides are identified to construct the torus.

Now, the construction of the physical states would a priori require to solve the Ashtekar-Barbero Hamiltonian constraint (\ref{F=0 in time gauge}). Just like in four-dimensional loop quantum gravity, it is not known wether or not at the physical level the $\gamma$ dependency will disappear and the nature of the geometric operators change. However, in the present three-dimensional model, the route to constructing the physical states is known. One can go back to the complex connection (\ref{decomposition of A}), impose the simplicity constraint that turns it into an $\su(1,1)$ connection, and then the Hamiltonian constraint becomes simply the flatness constraint of $\SU(1,1)$ BF theory. In this case, the scalar product between physical states is given by
\be
\langle\varphi_\text{phy}(s_1,s_2,s_3),\varphi_\text{phy}(s'_1,s'_2,s'_3)\rangle=\prod_{i=1}^3\delta(s_i-s_i'),
\ee
where $\delta(s-s')$ is the Dirac delta distribution, and any other quantization scheme (like the $\SU(2)$ formulation in the time gauge) should lead to the same result in order for it to be anomaly-free.

\section{Conclusion}

\noindent In this short paper, we have studied the loop quantization of three-dimensional gravity based on two different classical formulations of the Hamiltonian theory. Using the results of \cite{Imm3D}, we have first written the canonical theory in terms of $\SU(2)$ Ashtekar-Barbero variables with a real Barbero-Immirzi parameter. Then, we have argued that there is a simpler Hamiltonian formulation (in terms of a flatness constraint), in which the connection becomes complex and has to be supplemented by appropriate reality conditions. When these are implemented classically in the form of a linear simplicity constraint, one recovers an $\su(1,1)$ connection which does still depend on real values of the Barbero-Immirzi parameter $\gamma$. However, as we have shown, this connection leads to a quantum theory with $\gamma$-independent and continuous spectra for the geometric operators. It is important to stress out that what we did was not to compare the new $\SU(2)$ Ashtekar-Barbero formulation of three-dimensional gravity with usual first order Lorentzian gravity (i.e. $\SU(1,1)$ BF theory). Instead, starting from the $\SU(2)$ theory, we have naturally ended up with a new choice of connection that enables to work out the Hamiltonian constraint (just like the choice $\gamma=\pm\mathrm{i}$ in four dimensions), with the additional requirement that this connection should satisfy a reality condition similar to the spin foam linear simplicity constraint.

Which lessons can be learned for four-dimensional loop quantum gravity? The present model shows that already at the kinematical level one can have two very different quantum theories depending on the choice of connection. The same observation seems to be true in four dimensions, although it has never been made rigorous because of the absence of a kinematical arena for the self-dual connection of Ashtekar or the non-commutative shifted connection of Alexandrov. Moreover, it is clear in this three-dimensional model that physical states are $\SU(1,1)$ spin networks, therefore raising the question of wether the implementation of the Ashtekar-Barbero Hamiltonian constraint (\ref{F=0 in time gauge}) will also lead to these physical states, and to the disappearance of the Barbero-Immirzi parameter. Even if we have not addressed this question here, our model could be taken as a starting point to investigate this direction.

To summarize, if one tries to support the point of view that $\gamma$ should not play a role at the physical level, there are essentially three possible scenarios.
\begin{enumerate}
\item The first one is a quantum theory based on the $\SU(2)$ kinematics, and in which the Barbero-Immirzi parameter disappears in the quantum theory at the level of physical states (with possibly the spectra becoming continuous). From what is known today about the quantum scalar constraint, this does not seem to be the case. Moreover, the EPRL and FK${}_\gamma$ spin foam models are designed in such a way that $\gamma$ plays a central role in the definition of the dynamics of the theory.
\item Another possibility would be to go back to the self-dual connection already at the classical level, and to find a way of dealing with the reality conditions. This is what has been done in the three-dimensional model of \cite{Imm3D} and in the present paper. As we have seen, the reason for which this is possible in three-dimensions is that there is a natural notion of reality for the complex connection, which corresponds to considering its non-compact subgroup $\SU(1,1)$ by means of the linear simplicity constraint. Then, the non-compactness can be handled due to the topological nature of the theory, and the resulting possibility of working on a fixed minimal graph.
\item It seems to us that the most promising direction would be to consider $\gamma$ as a regulator needed in order to deal with the non-compactness of the gauge group. Because the four-dimensional theory has local degrees of freedom, this non-compactness cannot be handled like in the three-dimensional case by fixing a minimal graph. The introduction of the Barbero-Immirzi parameter could therefore be seen as a generalized Wick rotation, which has to be unfolded at the end of the day in order to obtain physical results. This procedure of going through a compact gauge group in order to define the theory, and then using tools of analytic continuation to return to the non-compact case, was put forward in the computation of black hole entropy in four \cite{FGNP} and three dimensions \cite{BTZ} (although in this last reference the regulator is the Chern-Simons level rather than the Barbero-Immirzi parameter).
\end{enumerate} 

If this last interpretation of the Barbero-Immirzi parameter turns out to be the correct one, then one should reconsider the results derived from the kinematical structure of the $\SU(2)$ theory, since these rely heavily on the compactness of the gauge group and the value $\gamma\in\mathbb{R}$. The key open question is in fact that of knowing at which step to perform the analytic continuation back to the self-dual value $\gamma=\mathrm{i}$, if this can be given any well-defined meaning in the (canonical or covariant) full theory at all.

\section*{Aknowledgements}

\noindent MG is supported by the NSF Grant PHY-1205388 and the Eberly research funds of The Pennsylvania State University.


\begin{thebibliography}{99}

\bibitem{GN1} M. Geiller and K. Noui,
Testing the imposition of the spin foam simplicity constraints,
Class. Quant. Grav. \textbf{29} 135008 (2012), \texttt{arXiv:1112.1965 [gr-qc]}.

\bibitem{GN2} M. Geiller and K. Noui,
A note on the Holst action, the time gauge, and the Barbero-Immirzi parameter,
Gen. Rel. Grav. (2012), \texttt{arXiv.12125064 [gr-qc]}.

\bibitem{Imm3D} J. Ben Achour, M. Geiller, K. Noui and C. Yu,
Testing the role of the Barbero-Immirzi parameter and the choice of connection in loop quantum gravity,
(2013)\texttt{arXiv. [gr-qc]}.

\bibitem{alexandrov5} S. Alexandrov,
On choice of connection in loop quantum gravity,
Phys. Rev. \textbf{D 65} 024011 (2001), \texttt{arXiv:gr-qc/0107071}.

\bibitem{LQC} A. Ashtekar and P. Singh,
Loop quantum cosmology: A status report
Class. Quant. Grav. \textbf{28} 213001 (2011), \texttt{arXiv:1108.0893 [gr-qc]}.

\bibitem{BHentropy1} C. Rovelli,
Black hole entropy from loop quantum gravity,
Phys. Rev. Lett. \textbf{77} 3288 (1996), \texttt{arXiv:gr-qc/9603063}.

\bibitem{BHentropy2} A. Ashtekar, J. Baez and K. Krasnov,
Quantum geometry of isolated horizons and black hole entropy,
Adv. Theor. Math. Phys. \textbf{4} 1 (2000), \texttt{arXiv:gr-qc/0005126}.

\bibitem{BHentropy3} K. A. Meissner,
Black hole entropy in loop quantum gravity,
Class. Quant. Grav. \textbf{21} 5245 (2004), \texttt{arXiv:gr-qc/0407052}.

\bibitem{BHentropy4} I. Agullo, J. F. Barbero, J. Diaz-Polo, E. Fernandez-Borja and E. J. S. Villase\~nor,
Black hole state counting in loop quantum gravity: A number theoretical approach,
Phys. Rev. Lett. \textbf{100} 211301 (2008), \texttt{arXiv:gr-qc/0005126}.

\bibitem{FGNP} E. Frodden, M. Geiller, K. Noui and A. Perez,
Black hole entropy from complex Ashtekar variables,
(2012), \texttt{arXiv:1212.4060 [gr-qc]}.

\bibitem{BST} N. Bodendorfer, A. Stottmeister and A. Thurn,
Loop quantum gravity without the Hamiltonian constraint
Class. Quant. Grav. \textbf{30} 082001 (2013), \texttt{arXiv:1203.6525 [gr-qc]}.

\bibitem{pranzetti} D. Pranzetti,
Black hole entropy from KMS-states of quantum isolated horizons,
(2013), \texttt{arXiv:1305.6714 [gr-qc]}.

\bibitem{BN} N. Bodendorfer and Y. Neiman,
Imaginary action, spinfoam asymptotics and the 'transplanckian' regime of loop quantum gravity,
(2013), \texttt{arXiv:1303.4752 [gr-qc]}.

\bibitem{Freidel} L. Freidel,
A Ponzano-Regge model of Lorentzian 3-dimensional gravity,
Nucl. Phys. Proc. Suppl. \textbf{88} 237 (2000), \texttt{arXiv:gr-qc/0102098}. 

\bibitem{NP1} K. Noui and A. Perez,
Three dimensional loop quantum gravity: Physical scalar product and spin foam models,
Class. Quant. Grav. \textbf{22} 1739 (2005), \texttt{arXiv:gr-qc/0402110}.

\bibitem{NP2} K. Noui and A. Perez,
Three dimensional loop quantum gravity: Coupling to point particles,
Class. Quant. Grav. \textbf{22} 4489 (2005), \texttt{arXiv:gr-qc/0402111}.

\bibitem{rovelli-smolin-area-volume} C. Rovelli and L. Smolin,
Discreteness of area and volume in quantum gravity,
Nucl. Phys. \textbf{B 442} 593 (1995), \texttt{arXiv:gr-qc/9411005}.

\bibitem{ashtekar-lewandowski-volume} A. Ashtekar and J. Lewandowski,
Quantum theory of geometry II: Volume operators,
Adv. Theor. Math. Phys. \textbf{1} 388 (1998), \texttt{arXiv:gr-qc/9711031}.

\bibitem{giesel-thiemann-vol1} K. Giesel and T. Thiemann,
Consistency check on volume and triad operator quantisation in loop quantum gravity I,
Class. Quant. Grav. \textbf{23} 5667 (2006), \texttt{arXiv:gr-qc/0507036}.

\bibitem{giesel-thiemann-vol2} K. Giesel and T. Thiemann,
Consistency check on volume and triad operator quantisation in loop quantum gravity II,
Class. Quant. Grav. \textbf{23} 5693 (2006), \texttt{arXiv:gr-qc/0507037}.

\bibitem{thiemann-brunnemann} T. Thiemann and J. Brunnemann,
Simplification of the spectral analysis of the volume operator in loop quantum gravity,
Class. Quant. Grav. \textbf{23} 1289 (2006), \texttt{arXiv:gr-qc/0405060}.

\bibitem{brunnemann-rideout1} J. Brunnemann and D. Rideout,
Properties of the volume operator in loop quantum gravity I: Results,
Class. Quant. Grav. \textbf{25} 065001 (2008), \texttt{arXiv:0706.0469 [gr-qc]}.

\bibitem{brunnemann-rideout2} J. Brunnemann and D. Rideout,
Properties of the volume operator in loop quantum gravity II: Detailed presentation,
Class. Quant. Grav. \textbf{25} 065002 (2008), \texttt{arXiv:0706.0382 [gr-qc]}.

\bibitem{freidel-livine-rovelli} L. Freidel, E. R. Livine and C. Rovelli,
Spectra of length and area in 2+1 Lorentzian loop quantum gravity,
Class. Quant. Grav. \textbf{20} 1463 (2003), \texttt{arXiv:gr-qc/0212077}.

\bibitem{BTZ} E. Frodden, M. Geiller, K. Noui and A. Perez,
Statistical entropy of a BTZ black hole from loop quantum gravity,
JHEP \textbf{5} 139 (2013), \texttt{arXiv:1212.4473 [gr-qc]}.

\end{thebibliography}
\end{document}